\documentclass{elsart}
\usepackage[centertags]{amsmath}
\usepackage{amssymb}
\usepackage{graphicx}
\journal{Physics Letters B}
\date{19 November 2003}

\begin{document}

\begin{frontmatter}

\title{Lattice measurement of the rescaling of the scalar condensate }

\author[infnba,uniba]{P. Cea},
\author[infnct]{M. Consoli},
and \author[infnba]{L. Cosmai\corauthref{cor1}}
\address[infnba]{INFN - Sezione di Bari, I-70126 Bari, Italy}
\address[infnct]{INFN - Sezione di Catania, I-70126 Catania, Italy}
\address[uniba]{Dipartimento di Fisica, Univ. Bari, I-70126 Bari, Italy}
\corauth[cor1]{email: Leonardo.Cosmai@ba.infn.it}

\begin{abstract}
We have determined the rescaling of the scalar condensate 
$Z\equiv Z_\varphi$ near the critical line of a 4D Ising model. 
Our lattice data, supporting previous numerical indications, 
confirm the behaviour $Z_\varphi\sim \ln ({\rm cutoff})$. This result is
predicted in an alternative description of symmetry breaking where
there are no upper bounds on the Higgs boson mass from `triviality'. 
\vspace{1pc}
\end{abstract}

\begin{keyword}
lattice\sep spontaneous symmetry breaking \sep Higgs
\PACS 14.80.Bn, 11.10.-z, 11.15.Ha

\end{keyword}
\end{frontmatter}

There are many computational and analytical evidences pointing towards the
'triviality' of $\Phi^4$ theories in $3+1$ 
dimensions~\cite{Sokal_book,Lang:1993sy},
though a rigorous proof is still lacking.
Nevertheless these theories continue to be useful
and play an important role for unified model of electroweak interactions.
The conventional view, when used in the Standard Model, leads to predict a 
proportionality relationship between the squared Higgs boson mass $m_H^2$
and the known weak scale $v_R$ (246 GeV) through the renormalized 
scalar self-coupling
$g_R \sim 1/{\ln \Lambda}$. In this picture, where 
$m^2_H \sim g_R v^2_R$, the ratio $m_H^2/v_R^2$ is a cutoff dependent quantity
that becomes smaller and smaller when $\Lambda$ is made larger and larger.

This usual interpretation of triviality has important phenomenological 
implications. For instance, a precise measurement
of $m_H$, say $m_H=760 \pm 21$~GeV,  would constrain the 
cutoff $\Lambda$ to be smaller than 2~TeV leading to predict
the existence of 'new physics' at that energy scale.

On the other hand there is another possible interpretation of triviality, 
suggested in a series of 
papers~\cite{Consoli:1994jr,Consoli:1997ra,Consoli:1999ni}.
In this alternative approach triviality and spontaneous symmetry breaking 
can coexist for arbitrarily large values of the cutoff $\Lambda$.
The essential point is that the
`Higgs condensate' and its quantum fluctuations
undergo different rescalings when changing the ultraviolet cutoff. 
Therefore, the relation between $m_H$ and the physical $v_R$ is not
the same as in perturbation theory. 

To remind the basic issue we observe that, beyond perturbation theory, in
a broken-symmetry phase, there are {\it two} different definitions of
the field rescaling: a rescaling of the `condensate', say
$Z\equiv Z_\varphi$, and a rescaling of the fluctuations, say
$Z\equiv Z_{\text{prop}}$. To this end, let us
consider a one-component scalar theory and introduce the bare
expectation value $v_B=\langle\Phi_{\text{ latt}}\rangle$
associated with the `lattice' field as defined at the cutoff
scale. By $Z\equiv Z_\varphi$ we mean the rescaling that is needed
to obtain the physical vacuum field $v_R= v_B / \sqrt{Z_\varphi}$.
By `physical' we mean that the second derivative of the effective 
potential $V''_{\rm eff}(\varphi_R)$, 
evaluated at the rescaled field 
$\varphi_R= \pm v_R$, is precisely given by $m^2_H$. 
Since the second derivative of the effective potential 
$V''_{\rm eff}(\varphi_B)$, evaluated at the bare field $\varphi_B= \pm v_B$, 
is the bare
zero-four-momentum two-point function, this standard definition is
equivalent to define $Z_\varphi$ as:
\begin{equation}
\label{z1phi} 
Z_\varphi= m^2_H \chi_2(0)  \,,
\end{equation}
where $\chi_2(0)$ is the bare zero-momentum susceptibility.
On the other hand, $Z\equiv Z_{\text{prop}}$ is determined from
the residue of the connected propagator on its mass shell.
Assuming `triviality' and the K\'allen-Lehmann representation for
the shifted quantum field, one predicts $Z_{\text{prop}} \to 1$
when approaching the continuum theory. 

Now, in the standard approach
one  assumes  $Z_\varphi=Z_{\text {prop}}$ (up to small 
perturbative corrections). However, in
the different interpretation of triviality of
Refs.~\cite{Consoli:1994jr,Consoli:1997ra,Consoli:1999ni}
although $Z_{\text
{prop}}\to 1$, as in leading-order perturbation theory,
$Z_\varphi\sim \ln \Lambda $ is fully non perturbative and
diverges in the continuum limit. In this case, 
in order to obtain $v_R$ from the
bare $v_B$ one has to apply a non-trivial correction. As a result, 
$m_H$ and $v_R$ now scale uniformly in the continuum limit, and
the ratio $C=m_H/v_R$ becomes a cutoff-independent quantity. 

To check this alternative picture against the generally accepted
point of view, one can run numerical simulations of the theory and compare
the scaling properties of $Z=Z_\varphi$ with those of $Z=Z_{\text{ prop}}$. 
If the standard interpretation is correct, the lattice data for $Z_\varphi$
should unambiguosly approach unity when taking the continuum limit. 

In this respect, we observe that numerical evidence for different
cutoff dependencies of $Z_\varphi$ and $ Z_{\text{prop}}$ has
already been 
reported~\cite{Cea:1998hy,Cea:1999kn,Cea:1999zu}. In those
calculations, 
one was fitting the lattice data for the connected
propagator to the (lattice version of the) two-parameter form
\begin{equation}
\label{gprop} G_{\text{fit}}(p)= \frac{Z_{\text{prop}}}{ p^2 +
m^2_{\text{latt}} }.
\end{equation}
After computing the zero-momentum susceptibility
$\chi_{\text{latt}}$, it was possible to compare the value of
$Z_\varphi \equiv m^2_{\text{latt}} \chi_{\text{latt}}$
with the fitted $Z_{\text{prop}}$,
both in the symmetric and broken phases. While no difference was
found in the symmetric phase, $Z_\varphi$  and $Z_{\text{prop}}$
were found to be sizeably different in the broken phase. In fact, 
$Z_{\text{prop}}$ was very slowly varying and steadily approaching
unity from below in the continuum limit. On the other hand,
$Z_{\varphi}$ was found to rapidly increase {\it above} unity in
the same limit consistently with the logarithmic trend 
$Z_\varphi \sim \ln \Lambda$ predicted in 
Refs.~\cite{Consoli:1994jr,Consoli:1997ra,Consoli:1999ni}.

\begin{table}[t]
\label{table1}
\caption{
We compare our determinations of
$\langle |\phi| \rangle$ and $\chi_{\text{latt}}$ for given
$\kappa$ with corresponding determinations found in the
literature~\cite{Jansen:1989cw}.  In the algorithm column, 'S-W'
stands for the  Swendsen-Wang algorithm, while 'W' stands for the
Wolff algorithm.}
\begin{center}
\begin{tabular}{cccccc}
$\kappa$   &lattice  &algorithm
& $\langle |\phi| \rangle$  &$\chi_{\text{latt}}$  \\
\hline
0.074  &$20^3 \times 24$  &W                          &            &142.21 (1.11)   \\
0.074  &$20^3 \times 24$  &Ref.\cite{Montvay:1987us}  &            &142.6 (8)
\\ \hline
0.077  &$32^4$  &S-W    & 0.38951(1) &18.21(4)   \\
0.077  &$16^4$  &Ref.\cite{Jansen:1989cw}  & 0.38947(2) &18.18(2)
\\ \hline
0.076  &$20^4$  &W   & 0.30165(8) &37.59(31)   \\
0.076  &$20^4$  &Ref.\cite{Jansen:1989cw}    & 0.30158(2) &37.85(6)   \\
\end{tabular}
\end{center}
\end{table}

A possible objection to this strategy is that the two-parameter
form  Eq.~(\ref{gprop}), although providing a good description of
the lattice data, neglects higher-order corrections to the
structure of the propagator. As a consequence, one might object
that the extraction of the various parameters is affected in an
uncontrolled way. For this reason, we have decided to change
strategy by performing a new set of lattice calculations. Rather
than studying the propagator, we have addressed the
model-independent lattice measurement of the susceptibility. In
this way, {\it assuming} the mass values from perturbation theory,
one can obtain a precise determination of $Z_\varphi$ to be
compared with the perturbative predictions.

\begin{table}[t]
\label{table2}
\caption{
The details of the lattice simulations for each $\kappa$ corresponding to $m_{\text{input}}$.
In the algorithm column, 'S-W' stands for the Swendsen-Wang algorithm~\cite{Swendsen:1987ce},
while 'W' stands for the Wolff algorithm~\cite{Wolff:1989uh}. 'Ksweeps' stands for
sweeps multiplied by $10^3$.}
\begin{center}
\begin{tabular}{cccccc}
$m_{\text{input}}$   &$\kappa$   &lattice  &algorithm &Ksweeps
&$\chi_{\text{latt}}$  \\
\hline
0.4     &0.0759     &$32^4$  &S-W  & 1750  &41.714 (0.132)  \\
0.4     &0.0759     &$48^4$  &W    &   60  &41.948 (0.927)  \\ \hline
0.35    &0.075628   &$48^4$  &W    &  130  &58.699 (0.420)  \\ \hline
0.3     &0.0754     &$32^4$  &S-W  &  345  &87.449 (0.758)  \\
0.3     &0.0754     &$48^4$  &W    &  406  &87.821 (0.555)  \\ \hline
0.275   &0.075313   &$48^4$  &W    &   53  &104.156 (1.305) \\ \hline
0.25    &0.075231   &$60^4$  &W    &   42  &130.798 (1.369)  \\ \hline
0.2     &0.0751     &$48^4$  &W    &   27  &203.828 (3.058)  \\
0.2     &0.0751     &$52^4$  &W    &   48  &201.191 (6.140)  \\
0.2     &0.0751     &$60^4$  &W    &    7  &202.398 (8.614)  \\ \hline
0.15    &0.074968   &$68^4$  &W    &   25  &460.199 (4.884)  \\ \hline
0.1     &0.0749     &$68^4$  &W    &   24  &1125.444 (36.365)  \\
0.1     &0.0749     &$72^4$  &W    &    8  &1140.880 (39.025)  \\
\end{tabular}
\end{center}
\end{table}
Our numerical simulations  were performed in the Ising limit 
where a one-component $(\lambda\Phi^4)_4$ theory becomes
\begin{equation}
\label{ising}
S_{\text{Ising}} = -\kappa \sum_x\sum_{\mu} \left[ \phi(x+\hat
e_{\mu})\phi(x) + \phi(x-\hat e_{\mu})\phi(x) \right]
\end{equation}
and $\phi(x)$ takes only the values $\pm 1$ (in an infinite
lattice, the broken phase is found for $\kappa > 0.07475$). 
Using the Swendsen-Wang  and Wolff cluster algorithms we have computed
the bare
magnetization:
\begin{equation}
\label{baremagn}
 v_B=\langle |\phi| \rangle \quad , \quad \phi \equiv \sum_x
\phi(x)/L^4
\end{equation}
(where $\phi$  is the average field for each lattice configuration) and
the zero-momentum susceptibility:
\begin{equation}
\label{chi}
 \chi_{\text{latt}}=L^4 \left[ \left\langle |\phi|^2
\right\rangle - \left\langle |\phi| \right\rangle^2 \right] .
\end{equation}
We used different lattice
sizes at each value of $\kappa$ to have a check of the finite-size
effects. Statistical errors have been estimated using the
jackknife. Pseudo-random numbers have been generated using {\sc Ranlux}
algorithm~\cite{Luscher:1994dy,James:1994vv,Shchur:1998wi} with the highest
possible 'luxury'. 
As a check of the goodness of our simulations, we show in Table~1
the comparison with 
previous determinations of $\langle |\phi| \rangle$ and
$\chi_{\text {latt}}$ 
obtained by other authors
~\cite{Jansen:1989cw}).

As anticipated, rather than computing the Higgs mass on the lattice
we shall use the perturbative predictions for its value and adopt the 
L\"uscher-Weisz scheme~\cite{Luscher:1988ek}.
To this end, we shall denote by
$m_{\text{input}}$ the value of the parameter $m_R$ reported in
the first column of Table~3 in Ref.~\cite{Luscher:1988ek} for any
value of $\kappa$ (the { Ising limit corresponding to the value of
the other parameter $\bar{\lambda}=1$). 

Our data for $\chi_{\text {latt}}$
at various $\kappa$ are reported in Table~2
for the range
$0.1 \leq m_{\text{input}}\leq 0.4$
(the relevant $\kappa$'s for $m_{\text{input}}=0.15, 0.25, 0.275, 0.35$ have been
determined through a numerical interpolation of the data shown in the
L\"uscher-Weisz Table).
\begin{figure}[t]
\begin{center}
\includegraphics[width=0.8\textwidth,clip]{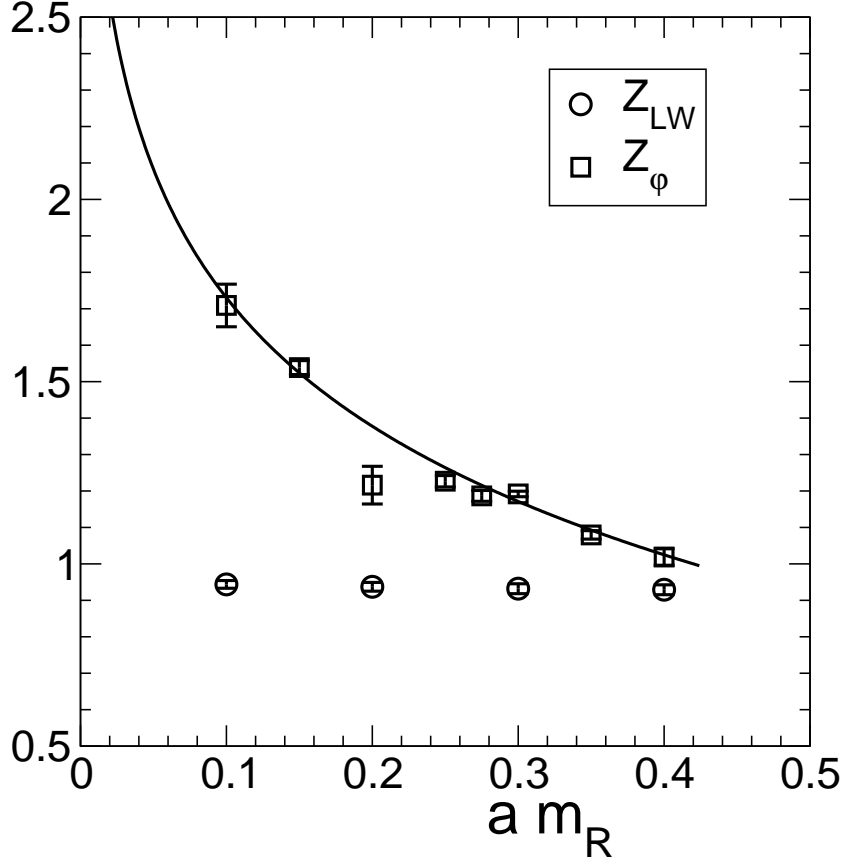}
\caption{The lattice data for $Z_\varphi$, as defined in
Eq.~(\ref{zphi}), and its perturbative prediction $Z_{\text{LW}}$
versus $m_{\text{input}}=a m_R$. The solid line is a fit to the form 
Eq.~(\ref{fit-form}) with $B=0.50$.}
\end{center}
\end{figure}
At this point, we can compare the quantity
\begin{equation}
\label{zphi}
       Z_\varphi\equiv 2\kappa m^2_{\text{input}} \chi_{\text{latt}}
\end{equation}
with the perturbative determination 
\begin{equation}
\label{zlw}
Z_{\text{LW}}\equiv 2\kappa Z_R
\end{equation}  
where $Z_R$ is defined in the third column of Table~3 in
Ref.~\cite{Luscher:1988ek}. 

The values of $Z_\varphi$ and 
$Z_{\text{LW}}$ are reported in Fig.~1.
We fitted the values for $Z_\varphi$ to the form ($\Lambda=\pi/a$)
\begin{equation}
\label{fit-form}
Z_\varphi= B \,  \ln{(\Lambda/m_R)} \,.
\end{equation}
As one can check, the two $Z$'s follow completely different trends
and the discrepancy becomes larger and
larger when approaching the continuum limit, precisely the same
trend found in 
Refs.\cite{Cea:1998hy,Cea:1999kn,Cea:1999zu}. This confirms
that, approaching the continuum limit, the rescaling of the `Higgs
condensate' cannot be described in perturbation theory.
Notice that the lattice data 
are completely consistent with the prediction 
$Z_\varphi \sim \ln \Lambda$ from
Refs.\cite{Consoli:1994jr,Consoli:1997ra,Consoli:1999ni}. 

On the other hand, for the symmetric phase, on the base of the theoretical 
predictions of
Refs.\cite{Consoli:1994jr,Consoli:1997ra,Consoli:1999ni} and on the
base of the numerical results of 
Refs.\cite{Cea:1998hy,Cea:1999kn,Cea:1999zu}, we do not expect deviations
from the perturbative predictions. As an additional 
check, we have computed the
zero-momentum susceptibility for the value 
$\kappa=0.0741$ that corresponds to 
$m_{\text{input}}=0.2$ (see Table~3 of Ref.~\cite{Luscher:1987ay}). From our 
value on a $32^4$ lattice $\chi_{\text{latt}}=161.94 \pm 0.67$,
using again Eq.~(\ref{zphi}), we obtain $Z_\varphi=0.960\pm 0.004$. When
compared with the corresponding L\"uscher-Weisz prediction
$Z_{\text{LW}}=0.975\pm 0.010$, this shows that, in the symmetric phase, 
lattice data and perturbation theory agree to good accuracy.

Let us now return to the broken phase. 
If the physical $v_R$ has to be computed from the bare $v_B$
 through $Z=Z_\varphi\sim \ln \Lambda$, rather than
through the perturbative $Z=Z_{\text{LW}}\sim 1$, one may wonder about the 
$m_H$-$v_R$ correlation. 
In this case the perturbative relation
\cite{Luscher:1988ek}
\begin{equation}
\label{gR} \left[ \frac{m_H}{v_R} \right]_{\text{LW}} \equiv \sqrt
{  \frac{g_R}{3} }.
\end{equation}
becomes 
\begin{equation}
\label{mh}
\frac{m_H}{v_R}= \sqrt{ \frac{g_R}{3}
\frac{Z_\varphi}{Z_{\text{LW}} } } \equiv C
\end{equation}
This is obtained by replacing $Z_{\text{LW}} \to Z_\varphi$ in
Ref.~\cite{Luscher:1988ek} but correcting for the perturbative
$Z_{\text{LW}}$  introduced in the L\"uscher and Weisz approach.
In this way, assuming the values of 
$g_R$ reported in the second column of
Table~3 of Ref.~\cite{Luscher:1988ek} and using our values of $Z_\varphi$
one gets a remarkably constant value of $C$. In fact, 
the $Z_\varphi \sim \ln \Lambda$ trend observed 
in Fig.1, compensates the $1/\ln \Lambda$ from $g_R$ so that
$C= 3.087\pm 0.084$ turns out to be a
cutoff-independent constant~\cite{Cea:2002zc,Cea:2003gp}.

A straightforward extension to the Standard Model
of this result leads to the cutoff independent value of 
the  Higgs boson mass $m_H=760\pm21$~GeV, corresponding to
$v_R=246$ GeV and to the Ising value $C=3.087 \pm 0.084$,
so that there are no upper bounds on $m_H$ from
`triviality'. In this sense, 
the whole issue of the upper limits on the
Higgs mass is affected suggesting the need of more extensive
studies of the critical line to compare the possible values of
$C$ in the full 2-parameter $\Phi^4_4$, both for the single-component and
four-component theory. 

In any case, a value as large as $m_H=760 \pm 21$~GeV would also be in good
agreement with a recent phenomenological analysis of radiative
corrections~\cite{Loinaz:2002ep} that points toward substantially
larger Higgs masses than previously obtained through global fits
to Standard Model observables.


\end{document}